\documentclass[
  ,numberedheadings 
  ]
  {aipproc}

\layoutstyle{6x9}

\begin{document}

\title{Coherent Phase Argument for Inflation}

\author{Scott Dodelson}{
  address={NASA/Fermilab Astrophysics Center,
Fermi National Accelerator Laboratory, Batavia, IL~~60510-0500\\
  Department of Astronomy \& Astrophysics, The University of Chicago,
Chicago, IL~~60637-1433} 
}

\begin{abstract}
Cosmologists have developed a phenomenally successful picture of structure in the universe 
based on the idea that the universe expanded exponentially in its
earliest moments.
There are three pieces of evidence for this exponential expansion -- {\it
inflation} -- from observations of anisotropies in 
the cosmic microwave background. First, the shape of the primordial spectrum is very similar
to that predicted by generic inflation models. Second, the angular scale at which
the first acoustic peak appears is consistent with the flat universe predicted by inflation.
Here I describe the third piece of evidence, perhaps the most convincing of all: the 
phase coherence needed to account for the clear peak/trough structure observed by the
WMAP satellite and its predecessors. I also discuss alternatives to inflation
that have been proposed recently and explain how they produce coherent phases. 
\end{abstract}

\maketitle


\section{Overview}

\begin{figure}[h,t]
  \includegraphics[height=.4\textheight]{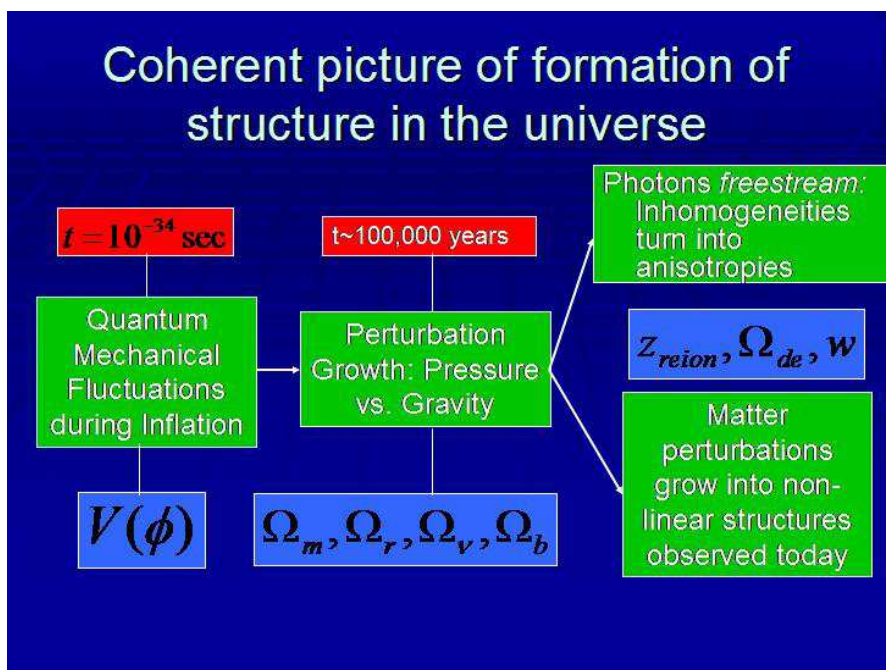}
  \caption{Outline of the evolution of structure in the universe. Perturbations
  are generated at very early times during inflation (determined by the potential
  $V$ of the field $\phi$ which drives inflation), start to evolve under the combined
  influence of pressure and gravity when the universe is of order $10^5$ years old,
  and then bifurcate into inhomogeneities in matter (which continue to grow due to
  gravity) and anisotropies in the radiation (which remain constant).}
  \label{fig:out}
\end{figure}

Over the last several years, we have gradually been accumulating
evidence for a remarkable theory of the early universe. This theory
now accounts for the observed structure in the universe by invoking 
new fundamental physics at very high energy scales. The theory is so elegant and
simple that it contains just a handful of free parameters. It is outlined in
Figure~\ref{fig:out}, which indicates how perturbations generated during inflation
evolve with time. The observations today of anisotropies in the radiation
and inhomogeneities in the matter therefore bear the imprint of: 
\begin{itemize}
\item the potential
  of the field(s) which drive(s) inflation
  \item the abundances of
  different types of matter in the universe
  (baryons, which interact with radiation; dark matter, which does not; and neutrinos,
  which can freestream out of overdense region): $\Omega_b,\Omega_m$, and $\Omega_\nu$
  \item late time phenomena such as dark energy (parametrized by abundance $\Omega_{de}$
  and equation of state $w$) and reionization
 \end{itemize}

The most important observations
confirming this model come from two segments of the electromagnetic spectrum. First, 
radio receivers have measured
the cosmic microwave background (CMB) to exquisite precision. Second, optical telescopes
have captured light from distant galaxies and quasars which tell us about 
the matter distribution both around those objects~\cite{2df,sdss} and along the line
of sight to us~\cite{weaklensing}. They have also received light from distant
objects such as galaxies and supernovae, allowing us to measure distances and fill in
a modern Hubble diagram~\cite{wendy,sn}.

Here I will focus on one observation and one aspect of the model. Anisotropies in
the CMB were first detected in 1992 by the COBE satellite~\cite{cobe}; were probed extensively
in the ensuing decade by more than thirty smaller scale experiments~\cite{smallscale}; and have now been
mapped exquisitely by the WMAP satellite~\cite{bennett}. These observations have been celebrated for 
measuring cosmological parameters to unprecedented accuracy~(e.g. \cite{spergel,mc}), and 
I will briefly describe
the progress on this front in \S 5. But, for the most part, I want to focus on how the signal
seen in the CMB is smoking gun evidence for the theory of inflation.

\section{Inflation}

The theory of inflation was introduced over twenty years ago~\cite{guth} to solve some of the
problems of the classical cosmology. For many years, progress was limited to
theory and to addressing the question of whether the density in the universe is indeed
equal to the critical density as inflation seems to predict. By far the most 
important, or at least the most testable, aspect of inflation though is its mechanism for
producing small perturbations about a smooth background. These perturbations can be measured as
long as we account for their (straightforward) evolution after inflation ends. 

It is well known that inflation produces perturbations characterized by a Harrison-Zel'dovich
spectrum~\cite{mc,inflper}. This means that the amplitude of a particular Fourier mode is drawn from
a distribution with mean equal to zero and variance given by
\begin{equation}
\langle \tilde\delta({\bf k}) \tilde\delta^*({\bf k'})\rangle
= (2\pi)^3 \delta^3({\bf k} - {\bf k'}) P(k)
\end{equation}
where $\delta$ is the fractional overdensity with power spectrum $P(k)$ proportional to $k^n$. A
Harrison-Zel'dovich spectrum corresponds to $n=1$, and most inflationary models predict something very close to
this. 
You might think then
that the shape of the power spectrum can be measured in observations, and this
is what convinces us that inflation is right. Well, it is true that we can measure the
power spectrum, both of the matter and of the radiation, and it is true that the observations
agree with the theory. But this is not what tingles our spines when we look at the data.

\begin{figure}
  \includegraphics[height=.3\textheight]{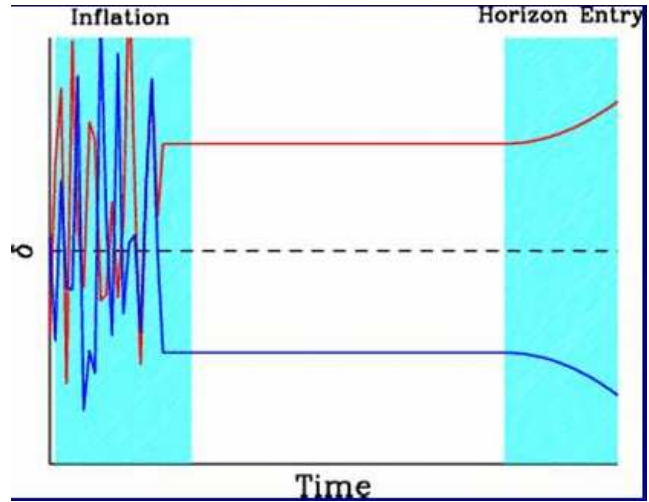}
  \caption{Evolution of the amplitudes of two Fourier modes with the same wavelength. After inflation, modes
  remain constant until they re-enter the horizon. After re-entry, they evolve under
  the competing influences of pressure and gravity.}
  \label{fig:inf}
\end{figure}

Rather, the truly striking aspect of perturbations generated during inflation is that all
Fourier modes
all have the same phase. To understand what this means and how it develops, 
consider Fig.~\ref{fig:inf} which shows a cartoon view of the evolution of the amplitudes of two
Fourier modes. Both oscillate quantum mechanically during inflation. Before inflation ends, though,
both {\it leave the horizon}, that is, their wavelengths get stretched so much that no causal physics
can alter them\footnote{Technically this occurs when the wavelength of the mode
becomes greater than the Hubble radius, $c/H$.}. Once this happens, their amplitudes remain constant. They stay constant up until the
time much later on (for modes of interest this might typically happen when the universe is 100,000
years old) when the modes re-enter the horizon, at which time causal physics can once
again affect their amplitudes. The crucial point here is that as the modes
re-enter the horizon, $\dot\delta$ is small. If we think of each Fourier mode as a linear
combination of a sin and a cos mode, inflation excites only the cos modes. It is difficult to envision
any other theory with this striking feature.

\section{Acoustic Oscillations}

How do perturbations evolve once they re-enter the horizon? A cartoon
version of the equation governing them is
\begin{equation}
\ddot\delta - c_s^2 \nabla^2\delta = F
\label{cs}
\end{equation}
where $c_s$ is the sound speed and $F$ is a forcing function due to gravity.
The perturbations obey the wave equation as one expects physically: a region
which is very overdense is driven by gravity to become more overdense, but
driven toward the average density by pressure. 

At this point, you might come to the conclusion that the spectrum of anisotropies
in the radiation today will exhibit a series of peaks and troughs just as a guitar string 
produces a series of higher harmonics. In fact, the spectrum of the CMB looks remarkably
like that of a guitar string. However, underlying the similarity is a pair of differences which
are essential to the argument that inflation is the origin of the perturbations.

A guitar string produces a set of harmonics because it is tied down at its ends. So there
are only a discrete set of frequencies at which it can oscillate. There is no such restriction
for perturbations in the early universe, so why do we see anisotropies at certain frequencies but
not at others? 

\begin{figure}
  \includegraphics[height=.3\textheight]{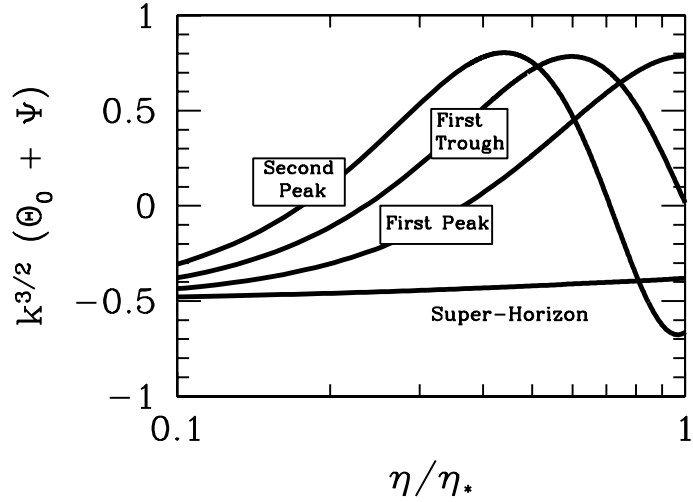}
  \caption{Evolution of four Fourier modes of the temperature of the radiation
  as a function of conformal time $\eta$ ($=\eta_*$ at recombination). Largest scale
  mode (labeled ``Super-Horizon'') is still constant at recombination. A slightly smaller scale mode (labeled ``First peak'')
  has begun its acoustic oscillation, and has maximal amplitude at recombination. An even smaller
  scale mode began oscillating earlier; its amplitude is zero at recombination. The smallest scale mode
  shown here (``Second Peak'') has gone through one full oscillation, so its amplitude will be at a
  maximum. From \cite{mc}.}\label{fig:overview}
\end{figure}

To understand this, consider Fig.~\ref{fig:overview} which shows the evolution of four Fourier
modes in the time leading up to recombination\footnote{After recombination, photons freestream
though the universe, so we see their distribution today as it was at the time of recombination.}.
The mode with the largest wavelength cannot be affected by causal physics so its amplitude remains
constant. Smaller scale modes have entered the horizon, and so have begun their acoustic oscillations.
The smaller the wavelength of a mode, the earlier it will have entered the horizon, and the more
oscillations it will have undergone by the time of recombination. Thus, the amplitude of the mode
labeled ``First Peak'' is maximal at recombination, and we expect to see large anisotropies on angular
scales which subtend this distance (roughly a degree). The mode labeled ``First Trough'' has oscillated
for a longer time though, and its amplitude is zero at recombination. Therefore, we expect very small
anisotropies on the corresponding angular scales. And on it goes, a succession of peaks and troughs
present not because no excitations are allowed at the frequencies in the troughs (as is the case for
the guitar string). Rather, perturbations are present at all wavelengths, but we happen to see only
some of them, depending on the phase of the oscillation at recombination.

\begin{figure}
 \hbox{
 \includegraphics[width=.45\textwidth]{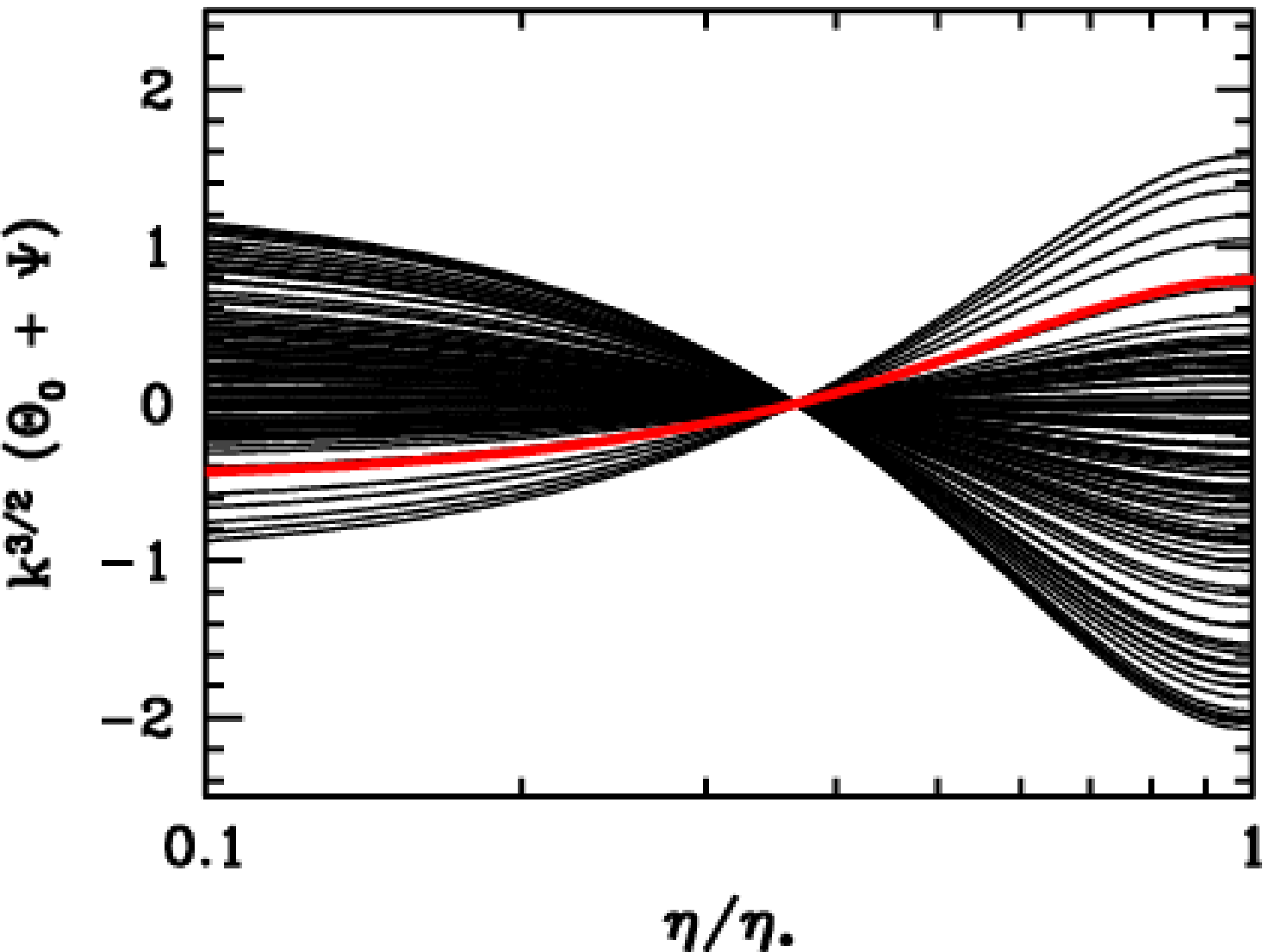}
 \hspace{9pt}
 \includegraphics[width=.45\textwidth]{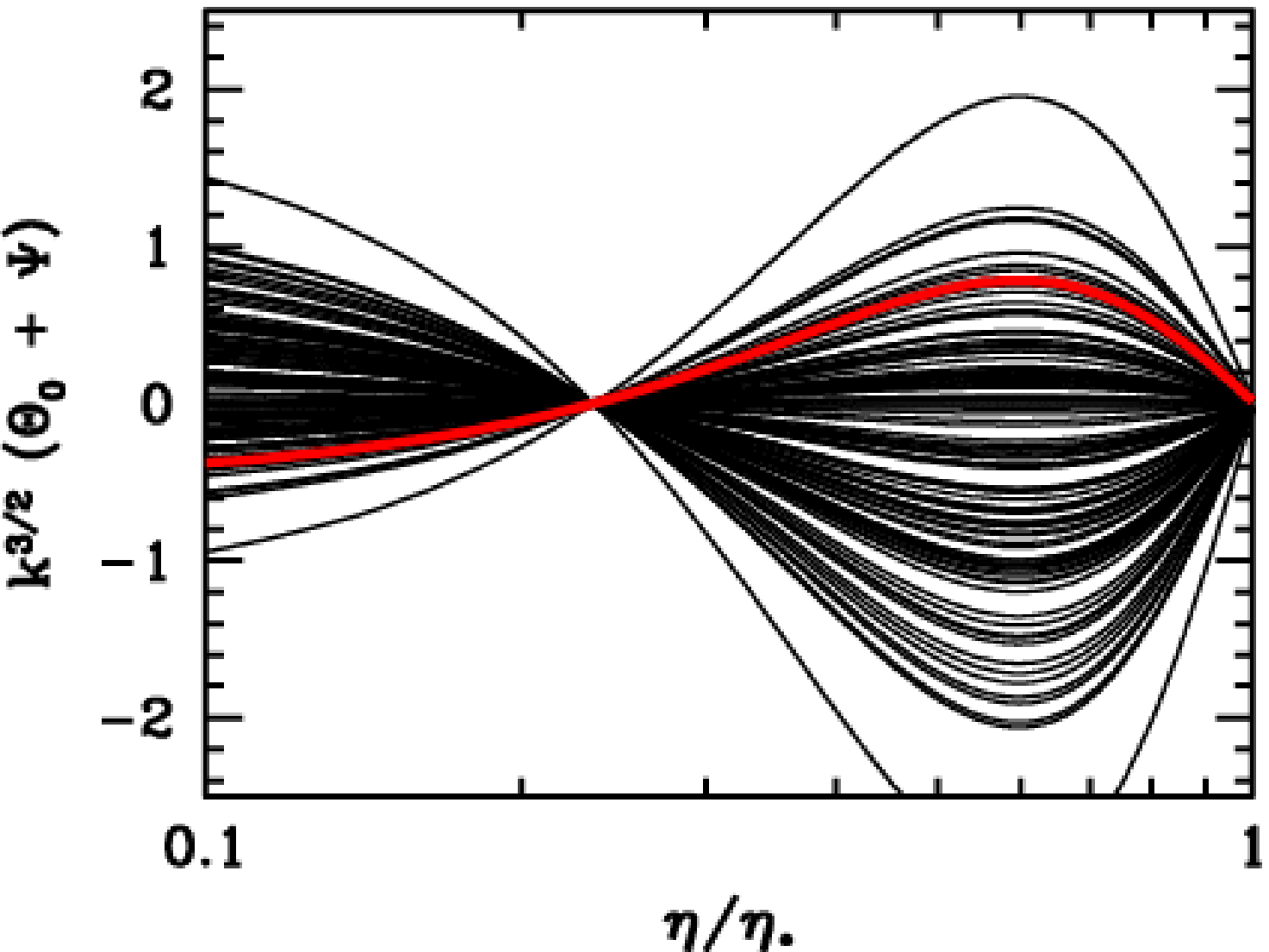}
  }
  \caption{The evolution of an infinite number of modes all with the same wavelength. Left panel shows
  the wavelength corresponding to the first peak, right to the first trough. Although the amplitudes of
  all these different modes differ from one another, since they start with the same phase, the ones on
  the left all reach maximum amplitude at recombination; the ones on the right all go to zero at
  recombination. }
  \label{fig:many}
\end{figure}

It is now very important to remember that there are many, many modes with nearly
identical wavenumbers. Think of the number of arrows that can point from the center of
a sphere to a fixed radius, keeping in mind that two arrows can be placed infinitesimally
close to each other. In fact, since the universe is effectively infinite, there are an
infinite number of modes for any wavenumber. All of these get excited during inflation and
we must sum over all of them to compute the anisotropy amplitude at a given scale. Thus, when 
I drew the single line corresponding to the ``First Peak'' mode in Fig.~\ref{fig:overview}, this was
really shorthand for an infinite number of modes all
with different amplitudes, as in Fig.~\ref{fig:many}. The amplitudes may differ, but as
Fig.~\ref{fig:many} shows, the phases are all the same. All modes enter the horizon with constant
amplitude. Thus, all modes with the ``First Peak'' wavenumber have maximal amplitude (left panel
in the figure) at recombination: they have all undergone half an oscillation, so their sign simply
changes. Similarly, all modes corresponding to ``First Trough'' have gone through $3/4$ of an
oscillation\footnote{You might expect the mode which has gone through $1/4$ of a full oscillation
to be the first trough. However, there are other effects (the dipole and the Integrated Sachs-Wolfe effect) which
fill in this trough.}; since they all are cosine modes and $\cos(3\pi/2)=0$, all have zero amplitude at
recombination (right panel).

\begin{figure}
 \hbox{
 \includegraphics[width=.45\textwidth]{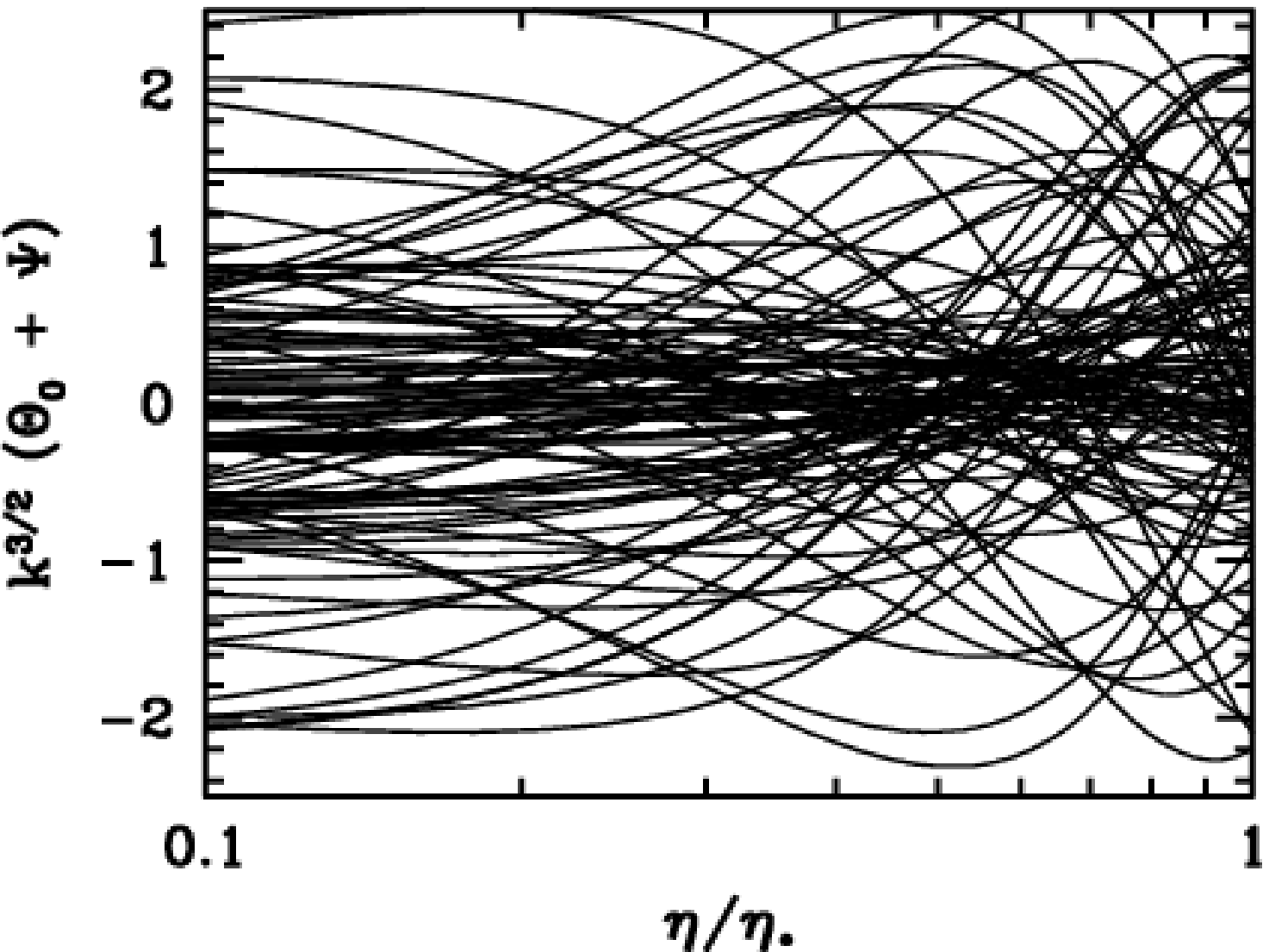}
 \hspace{9pt}
 \includegraphics[width=.45\textwidth]{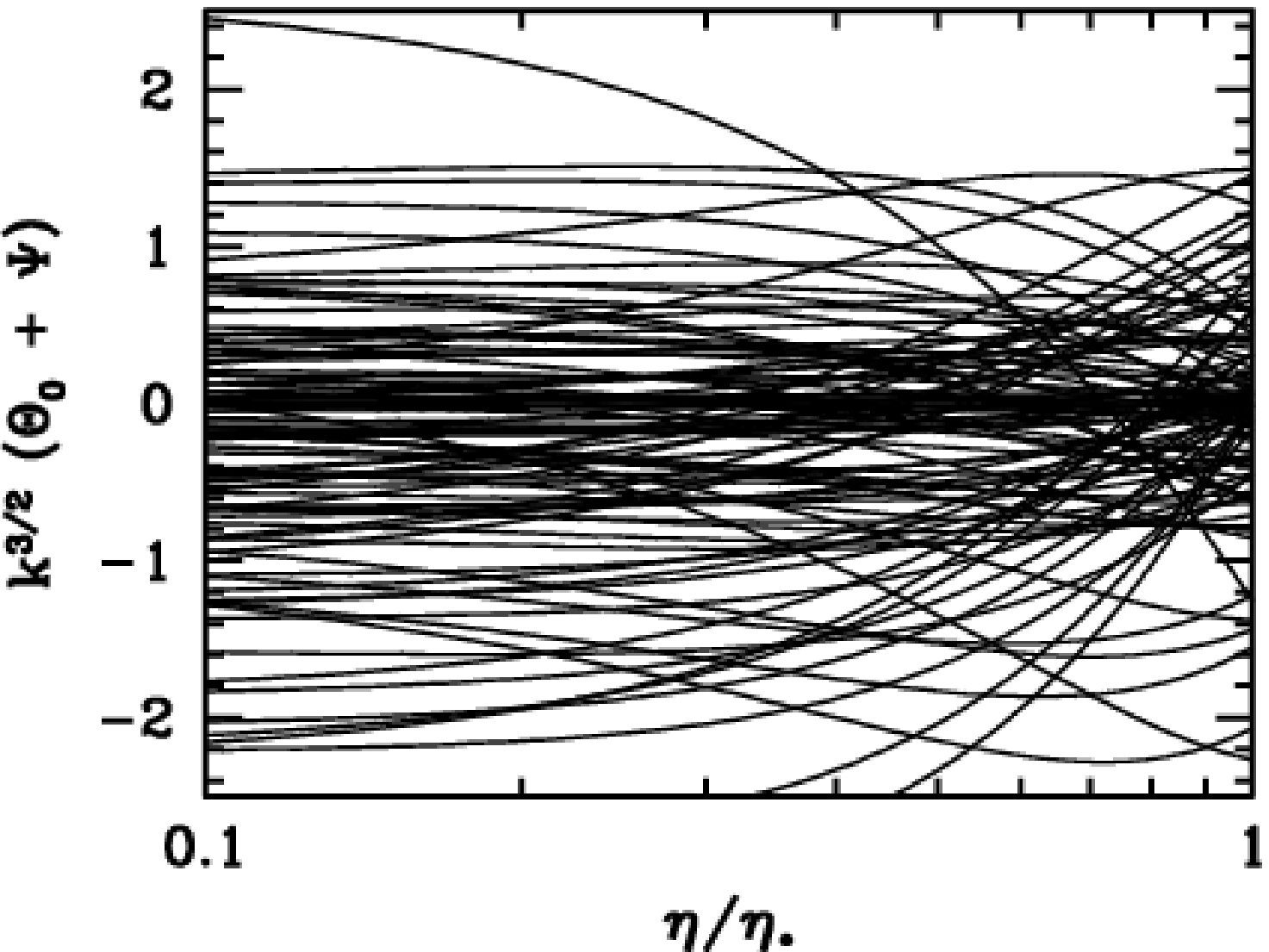}
  }
  \caption{Modes corresponding to the same two wavelengths ({\it First Peak} and {\it First Trough}) as in
  Fig.~\ref{fig:many}, but this time with initial phases scrambled. The anisotropies at the angular
  scales corresponding to these wavelengths would have identical rms's if the phases were random.}
\label{fig:dec}
\end{figure}

Contrast this with the situation one might otherwise expect, random phases, as depicted in
Fig.~\ref{fig:dec}. If the phases were truly random, so that both the sine and cosine modes were excited,
then at recombination, there would not be any difference at all between the rms amplitudes of the
{\it First Peak} and {\it First Trough} wavenumbers. So we would not see a sequence of peaks and
troughs in the anisotropy spectrum today. We would see simply a flat spectrum with no features.
If not for inflation, we {\it would} see a flat spectrum. How else could the phases have been
set well before the modes of interest entered the horizon?

\begin{figure}
  \includegraphics[height=.5\textheight]{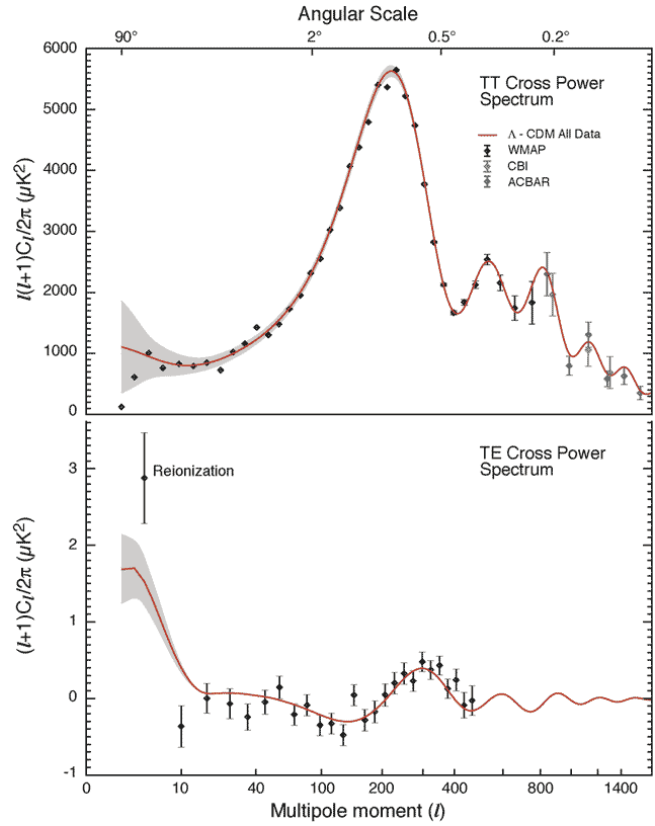}
  \caption{{\it Top panel:} Temperature anisotropies in the CMB as a function of angular scale~\cite{bennett}
  (smaller scales toward the right). The series of peaks and troughs is a clear indication of
  phase coherence, presumably coordinated during inflation. {\it Bottom panel:} Cross correlation
  between the temperature and the polarization as a function of angular scale. The anti-correlation
  at $100 < l < 200$ and positive correlation from $200 < l < 400$ are also due
  to phase coherence generated during inflation.}
  \label{fig:wmap}
\end{figure}

Therefore, when we look at the anisotropy spectrum recently measured by WMAP~\cite{bennett} and we
see the first and second peaks and troughs very clearly (Fig.~\ref{fig:wmap}), we are really observing
inflation doing the work of coordinating the phases of all Fourier modes. Without this coherence,
there would be no peaks and troughs.

\section{Polarization}

The bottom panel of Figure~\ref{fig:wmap} shows the cross-correlation between the temperature
and polarization anisotropies. This cross-correlation was first detected by the DASI experiment
in late 2002~\cite{kovac}, so our measurements of polarization are much less established than
those of temperature. Yet the WMAP results already are a crucial part of the coherence argument for
inflation. The peaks and troughs in the anisotropy spectrum all are on angular scales smaller than
a degree $(l>200)$; all of these scales were within the horizon at the time of recombination. So
you might imagine that one could concoct a theory of structure formation which obeyed causality and
still managed to produce only the cosine modes. If you could concoct such a theory~\cite{turok,hw},
then you could explain the peaks and troughs without appealing to inflation. It seems unlikely, but it
is at least logically possible. 

This logical possibility evaporates when confronted with the polarization data. In particular,
the negative cross-correlation between temperature and polarization on scales  $100 < l < 200$
is also the result of phase coherence, as we will shortly see, and the scales involved were {\it not}
within the horizon at recombination. So there is no causal mechanism that could have produced this
anti-correlation: we {\it must} appeal to inflation to understand it.

\begin{figure}
  \includegraphics[height=.8\textwidth]{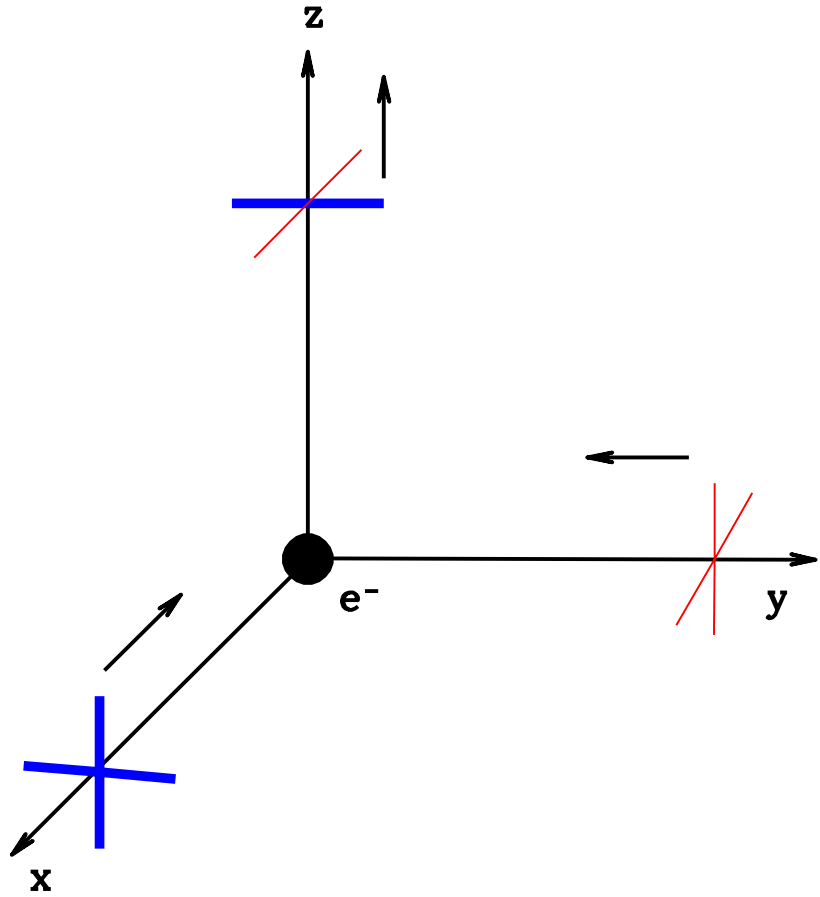}
  \caption{Incoming unpolarized
  radiation along the $x$- and $y$- axes produces outgoing polarized
  radiation along the $z$-axis only if the initial distribution has a non-zero 
  quadrupole moment (figure from \cite{mc}.)}\label{fig:quadray}
\end{figure}

To understand why the temperature and polarization are anti-correlated on scales of order
a degree, we first must establish that polarization results from Compton scattering of
a radiation field with a quadrupole moment. To see this consider Fig.~\ref{fig:quadray}
which depicts incoming radiation in the $z=0$ plane and shows the polarization
of the outgoing radiation along the positive $z$-axis. Since the radiation
has a quadrupole, incoming radiation along the $x$-axis is hotter than that along the $y$-
axis. Only the $y$-component of the polarization of the incoming $x$-ray gets transmitted
(the $z$-component is parallel to the outgoing direction, and polarization is transverse
to the direction of propagation) and only the $x$-component of the incoming $y$-ray
gets transmitted. Hence the outgoing $x$-component is cooler than the outgoing $y$-component.
A quadrupole in an unpolarized radiation field produces polarized radiation after Compton 
scattering. 

Therefore, the polarization today should be proportional to the quadrupole at the time of 
recombination. The photons just before recombination are tightly coupled to the electrons. 
This tight coupling tends to suppress the quadrupole. Consider an observer measuring incoming
photons, and for simplicity assume there is only a single plane wave perturbation. When the
observer looks perpendicular to the direction along which the density is varying, he sees
no perturbation. 
When he measures along this direction,
he measures a Doppler shift, which can be Taylor expanded as 
\begin{equation}
{\delta T\over T} = v + v'\, \lambda_{\rm mfp}
\end{equation}
where $v$ is the electron velocity; $v'$ its spatial derivative which is of order
$v/\lambda$ with $\lambda$ the wavelength of the perturbation; and $\lambda_{\rm mfp}$ the
distance the photon has traveled since it last scattered, the mean free path. 
The first term here represents the dipole seen
by our hypothetical observer, the second the quadrupole. Hence the quadrupole is
proportional to $v \lambda_{\rm mfp}/\lambda$. The quadrupole then
is proportional to the electron velocity. The dipole of the radiation is equal to
the electron velocity, so the quadrupole is proportional to the dipole right before
recombination. The proportionality constant is small, since $\lambda_{\rm mfp}$ is much smaller
than the typical wavelength. Collecting these arguments, we expect
\begin{equation}
P \simeq D {\lambda_{\rm mfp} \over \lambda}
\end{equation}
where $P$ is the polarization and $D$ the dipole. Incidentally, this explains why
measurements of polarization have lagged behind those of temperature anisotropies: the
polarization signal is a factor of ten smaller due to the $\lambda_{\rm mfp}/\lambda$
suppression.

\begin{figure}
  \includegraphics[height=.7\textwidth]{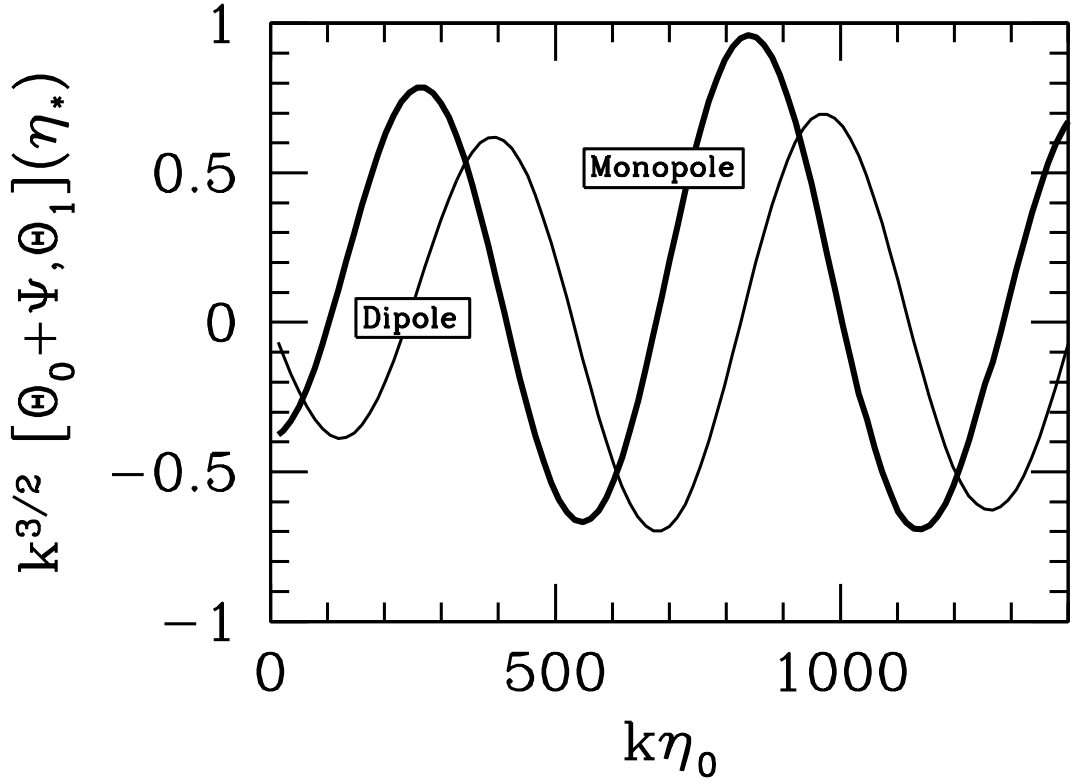}
  \caption{The monopole and dipole of the radiation field at recombination
  as a function of wavenumber $k$ (from \cite{mc}). A perturbation with wavenumber $k$ shows up
  on angular scales $l\sim k\eta_0$ where $\eta_0$ is (roughly) the distance to the
  last scattering surface (angle $\theta\sim l^{-1}$ is equal to size of object
  $k^{-1}$ divided by distance to last scattering surface).}
  \label{fig:dipdec}
\end{figure}

The polarization of the CMB today then is determined by the dipole at recombination. The temperature
anisotropies on the other hand are determined by the monopole at recombination\footnote{The {\it monopole} is what
normally thinks of when speaking of the temperature in a given spot. It is the average temperature of
all photons hitting that spot coming from all directions.}. The cross-correlation between the
temperature anisotropy and the polarization anisotropy then is proportional to the cross-correlation of
the monopole and dipole at recombination. How is the monopole related to the dipole at recombination? We can
answer this simply by recalling the continuity equation:
\begin{equation}
{\partial \rho\over \partial t} + \nabla \cdot (\rho {\bf v}) = 0
.\end{equation}
The velocity then (or equivalently the dipole) is proportional to $\dot\rho$, the time derivative of the
monopole. This is shown explicitly in Fig.~\ref{fig:dipdec}. At recombination, this phase difference
causes the product of the two to be negative for $100 < l < 200$ and positive on smaller scales until
$l\sim 400$. But this is precisely what WMAP has observed! We have clear evidence that monopole and
dipole were out of phase with each other at recombination. 

This evidence is exciting for the small scale modes ($l>200$). Just as the acoustic peaks
bear testimony to coherent phases, the cross-correlation of polarization and temperature
speaks to the coherence of the dipole as well. It solidifies our picture of the plasma
at recombination. The evidence from the larger scale modes ($l<200$) though is positively stupendous.
For, these modes were not within the horizon at recombination. So the {\it only} way they could have
their phases aligned is if some primordial mechanism did the job, when they were in causal contact.
Inflation is just such a mechanism.

\section{Cosmological Parameters}

\begin{figure}
  \includegraphics[height=.7\textwidth]{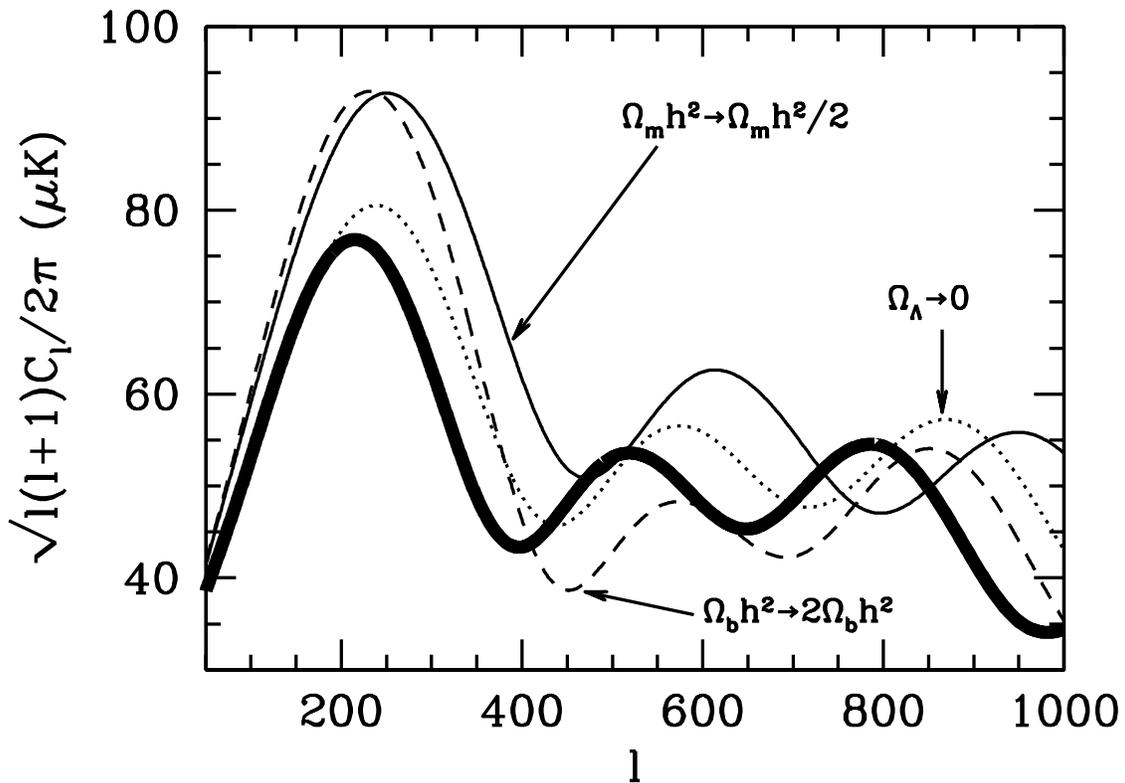}
  \caption{Dark solid curve is a model with $70\%$ cosmological constant and thirty percent baryons.
  Lighter curves show how the anisotropies change when varying different parameters. Here the
  total density is set to the critical density. From \cite{mc}.}
  \label{fig:hard}
\end{figure}

I hope I have convinced you that we now have very good reason to believe in the basic framework of
inflation as the seed of structure in the universe. Once we assume this framework, we can go ahead and
determine the free parameters in the model. The first and most renowned is that the first peak
appears where it should if the universe is flat, so let's assume that the universe is flat. Three of the
easiest parameters to measure then are the baryon density $\Omega_b$, the matter density $\Omega_m$,
and the Hubble constant $h$. Actually, as indicated in Fig.~\ref{fig:hard}, the CMB anisotropies
are most sensitive to combinations of these parameters. 

Here is a rough guide to the sensitivity of the anisotropies to these three cosmological parameters~\cite{mc}.
\begin{itemize}

\item {\it Baryon density} $\Omega_b h^2$. The sound speed in Eq.~\eqref{cs} goes down as more baryons
are added. The frequency of oscillation thus becomes smaller as the baryon density
goes up. A reduced frequency accentuates the effectiveness of the driving force, making the
oscillation more asymmetric. The result is that the height of the second peak is much smaller
than the height of the first peak when the baryon density is high.

\item {\it Matter density} $\Omega_m h^2$. If there is a lot of radiation at recombination, the
gravitational potential changes with time, inducing a larger driving force and hence boosting
anisotropies.

\item {\it Cosmological constant} $\Omega_\Lambda$. The cosmological constant\footnote{In a flat universe with constant $\Omega_m h^2$
and $\Omega_b h^2$, reducing $\Omega_\Lambda$ is equivalent to raising one of the other $\Omega$'s and
reducing the Hubble constant $h$ to keep the products $\Omega_ih^2$ fixed.} is
a late time effect, so the only impression it leaves on the CMB relates to the way physical scales
project onto angular scales; i.e. $\Omega_\Lambda$ changes the distance to the last scattering surface,
so the curves simply shift horizontally if $\Omega_\Lambda$ changes.

\end{itemize}

\begin{figure}
  \includegraphics[height=.8\textwidth]{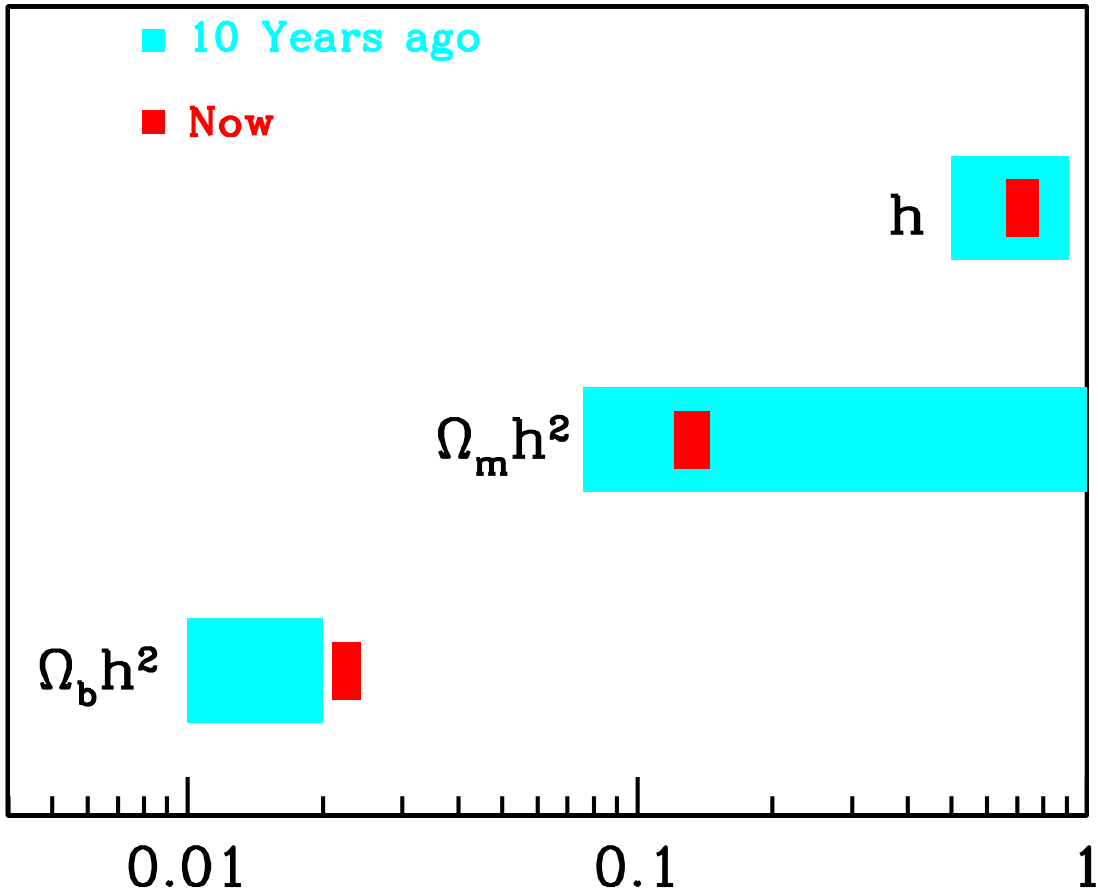}
  \caption{Allowed ranges for three cosmological parameters assuming the universe is flat. 
  Light hatched bands are a rough estimate of
  numbers used ten years ago. The dark solid bands are 1-sigma errors from WMAP~\cite{spergel}.}
  \label{fig:data}
\end{figure}

The results from WMAP~\cite{spergel} are shown in Fig.~\ref{fig:data}. We get a sense that the
CMB has reduced parameter uncertainties by close to a factor of ten. And this improvement allows us to make
several remarkable statements about our universe, based solely on observations of the CMB.
First, if one assumes the universe is 
exactly flat, then the CMB tells us that Hubble constant is $h=0.72\pm0.05$, in remarkable agreement
with direct determinations~\cite{wendy}. The ratio of the total matter density to the baryonic
density is about $6\pm1$, which means the CMB alone requires significant non-baryonic dark matter.
Finally, $\Omega_\Lambda=1-\Omega_m$ is equal to $0.71\pm 0.07$. The CMB, together with the flatness
assumption, requires a cosmological constant, or some form of dark energy. No wonder parameter
determination has received so much publicity!

\section{Alternatives}

It is perhaps not surprising that, as the evidence for inflation has firmed up,
theoreticians have been working harder than ever to find alternatives to
inflation. Here I want to focus on the question of what alternatives are viable
in light of the coherent phase argument. 

It is instructive to start with two models that don't quite make it.
The first is the well-known class of models with structure seeded by topological
defects. The phases of the Fourier modes are not synchronized in defect models,
so we do not expect a coherent series of peaks and troughs. This elegant
argument was first first advanced by Albrecht et al. \cite{andy} as a way of
distinguishing defect models from inflation and later confirmed in detailed
numerical studies~\cite{allen,pst}.

A second alternative has recently been proposed by Armendariz-Picon and
Lim~\cite{lim}. They note that inflation works by producing perturbations when
the modes of interest are sub-horizon and then driving these modes to be larger
than the horizon. Once outside the horizon, the
perturbations freeze-out, i.e. remain constant, until they re-enter the horizon
much later around the time of recombination. They point out that really perturbations freeze out once they
leave the {\it sound horizon 
horizon} ($c_s/H$ instead of $H^{-1}$). Thus instead of the Hubble rate
remaining roughly constant (as during inflation wherein $c_s=1$), freeze-out can also be
accomplished if the sound speed drops rapidly. This is a clever idea, one that
might ultimately be part of a viable alternative. At present though, it doesn't
quite work, because -- as they point out -- inflation is still needed (after
perturbation production) to drive the scale beyond the horizon\footnote{However, the
since scale-invariant perturbations have already been generated, the requirements on
inflation are less severe than usual.}. Another way of
saying this is to notice that 
if not for inflation, the modes of interest would never have been sub-horizon,
so nothing could have been effective in producing perturbations. 

Indeed, the most basic requirement for coherent phases is that at some point is
the distant past (well before recombination), the modes of interest had to be within 
the horizon. So, the coherent phases requirement is simply a strengthened
version of the classical horizon problem. With that in mind, I come to a final
class of alternatives which generate perturbations in a variety of ways, but
all share the same innovative approach to the horizon problem.

\begin{figure}
  \includegraphics[height=.7\textwidth]{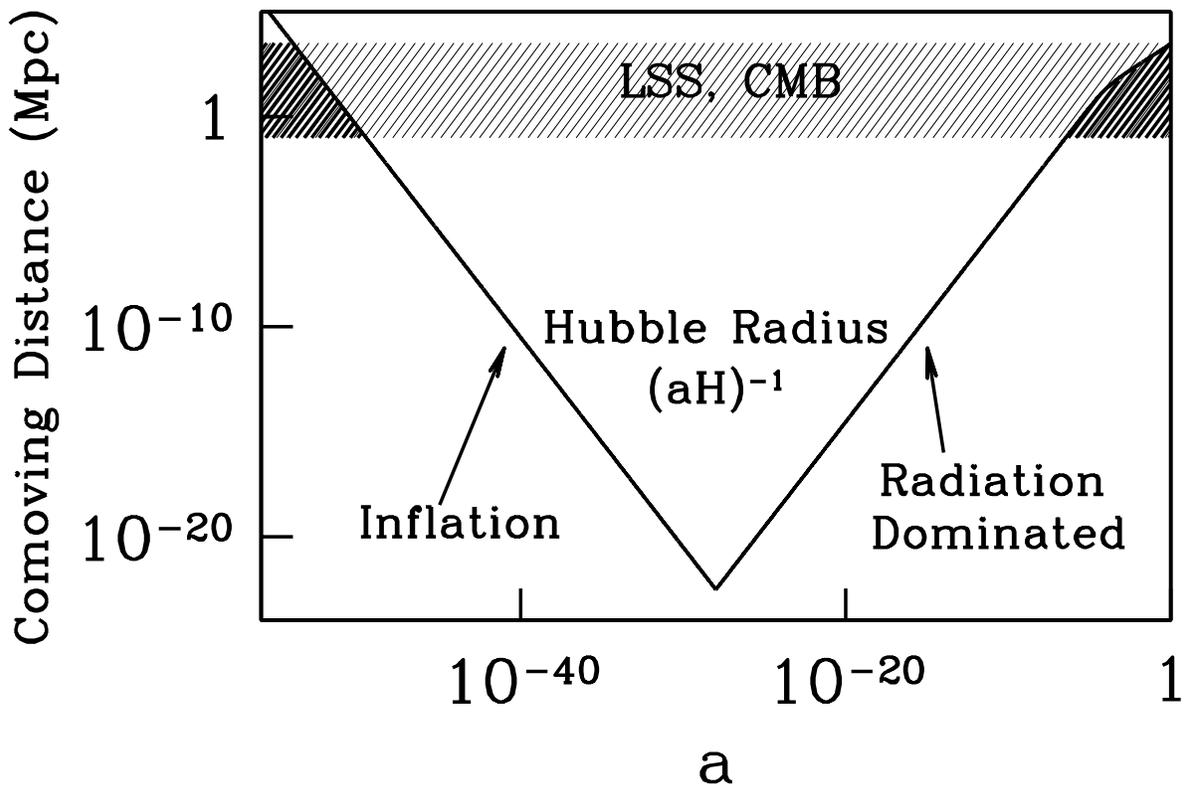}
  \caption{Evolution of the comoving Hubble radius in the inflationary picture.
  During inflation $H$ remains, constant, so the comoving Hubble radius drops as
  the universe expands exponentially. Thus scales which were initially in causal
  contact (at $a\sim 10^{-50}$ in the figure) freeze-out. From \cite{mc}.}
  \label{fig:hubble_radius}
\end{figure}

One way to think of the horizon problem is in term of the comoving grid, wherein
the wavenumber of any mode remains constant with time. The comoving Hubble
radius is $(aH)^{-1}$, which typically increases with time. The classical
horizon problem is that, since the comoving Hubble radius monotonically
increases in the standard cosmology, these modes must have all been outside the
horizon before recombination. The inflationary solution is depicted in
Fig.~\ref{fig:hubble_radius}, which shows that the necessary requirement is that
$aH$ must have {\it increased} sometime in the past.
Mathematically then we would seem to require
\begin{equation}
{d\over dt} \left( aH\right) = {d^2a \over dt^2} > 0
\end{equation}
at some point in the distant past. That is, we seem to require inflation (which
can be defined as a period in which $\ddot a > 0$).
The alternative models~\cite{alternate} however evade this constraint by using a contracting
phase. The requirement that the comoving radius decrease now is
$d[-aH]/dt > 0$ (since the Hubble radius is $-H^{-1}$) or $\ddot a < 0$.
So the horizon problem can be solved and the necessary coherent phases
generated if: (i) the universe accelerates while it is expanding (inflation)
OR (ii) the universe decelerates while it is contracting.

These alternatives and others are honing in on the question of what we have
really learned from the observations. That is, it is no longer sufficient to solve the
classical horizon problem. While the scales of interest are sub-horizon, a mechanism is
needed to generate perturbations with the proper amplitude and shape and to drive these
perturbations beyond the horizon so they freeze out.

\section{Conclusions}

Detailed observations of the CMB have solidified our confidence in a model of structure formation
based on inflation. The most striking evidence for inflation is the phase coherence of the primordial
perturbations, which manifests itself in the peaks and troughs of the temperature anisotropies and
in the cross-correlation between the temperature and the polarization. Once this framework has been accepted, it is
possible to use it and extract cosmological parameters. This parameter estimation suggests that: the universe is
flat; there is non-baryonic dark matter; and dark energy dominates the energy budget. 

As the observations
have improved, theorists have expanded the range of models which can account for them. and
proposed new alternatives to
inflation. The exciting development is that these alternatives 
must solve a much more demanding version of the horizon problem. And we are learning more
about what precisely is necessary to generate the coherent phases we have so unambiguously observed.


\begin{theacknowledgments}
   This work was supported by the DOE, by NASA Grant NAG5-10842, and by NSF
   Grant PHY-0079251. I would also like to thank the organizers of the Fourth Tropical
   Workshop on Particle Physics and Cosmology: R. Volkas, S. Tovey, C. N. Leung, and J. Nieves.
\end{theacknowledgments}





\end{document}